\documentclass[letter]{aa} 

\usepackage{txfonts}
\usepackage{longtable}
\usepackage{rotating}
\usepackage{natbib}
\usepackage{graphicx}
\usepackage{graphics}
\usepackage{psfrag}
\usepackage{amssymb}
\usepackage{xspace}
\usepackage{color}
\bibpunct{(}{)}{;}{a}{}{,}
\def\Teff{\ensuremath{T_{\mathrm{eff}}}}
\def\logg{\ensuremath{\log g}}

\def\vmic{$\upsilon_{\mathrm{mic}}$}
\def\vmac{$\upsilon_{\mathrm{macro}}$}
\def\vsini{\ensuremath{{\upsilon}\sin i}}

\def\kms{$\mathrm{km\,s}^{-1}$}

\def\logl{\ensuremath{\log L/L_{\odot}}}

\def\M{\ensuremath{M_{\odot}}}
\def\R{\ensuremath{R_{\odot}}}

\def\bz{$\langle$B$_z\rangle$}
\def\nz{$\langle$N$_z\rangle$}
\def\sbz{$\sigma_{\langle\,B_z\,\rangle}$}

\newcommand{\xmm}{{\em XMM-Newton}}
\newcommand{\chandra}{{\em Chandra}}
\newcommand{\rhoOph}{$\rho$~Oph}
\newcommand{\fxu}{{erg~s$^{-1}$~cm$^{-2}$}}
\newcommand{\lxu}{{erg~s$^{-1}$}}

\begin{document}
\title{Detection of magnetic field in the B2 star \rhoOph~A with ESO FORS2.
\thanks{Based on observations collected at the European Organisation for Astronomical Research in the Southern 
Hemisphere under ESO programmes 099.D-0067(A) and 078.C-0403(A).}}
\subtitle{}
\author{I. Pillitteri\inst{1,2}			\and
	L. Fossati\inst{3} 			\and
	N. Castro Rodriguez\inst{4}			\and
	L. Oskinova\inst{5,6}                   \and
        S. J. Wolk\inst{2}		
}
\institute{
	INAF -- Osservatorio Astronomico di Palermo G.S. Vaiana, Piazza del
	Parlamento 1, 90134, Palermo, Italy
	\email{pilli@astropa.inaf.it}
	\and
	Harvard-Smithsonian Center for Astrophysics, 60 Garden St., Cambridge,
	MA, USA
	\and
	Space Research Institute, Austrian Academy of Sciences, Schmiedlstrasse
	6, A-8042 Graz, Austria
	\and
	Department of Astronomy, University of Michigan, 1085 South University Avenue, 
	Ann Arbor, Michigan 48109, USA
	\and
	Institut f\"ur Physik und Astronomie, Universit\"at Potsdam
	Karl-Liebknecht-Strasse 24/25 14476 Potsdam-Golm, Germany
	\and
	Kazan Federal University, Kremlevskaya Str., 18, Kazan, Russia
}
\date{}
\abstract
{
Circumstantial evidence suggests that magnetism and enhanced X-ray emission are likely correlated in early B-type 
stars: similar fractions of them ($\sim$ 10\%) are strong and hard X-ray sources and possess strong 
magnetic fields. It is also known that some B-type stars have spots on their surface. 
Yet up to now no X-ray activity associated with spots on early-type stars was detected. 
In this Letter we report the detection of a magnetic field on the B2V star $\rho$ Oph A. 
Previously, we assessed that the X-ray activity of this star is associated with a surface spot, 
herewith we establish its magnetic origin. We analyzed FORS2 ESO VLT spectra of $\rho$ Oph A taken at 
two epochs and detected a longitudinal component of the magnetic field of order of $\sim500$ G in one of the datasets.  
The detection of the magnetic field only at one epoch can be explained by stellar rotation which is also 
invoked to explain observed periodic X-ray activity. From archival HARPS ESO VLT high resolution spectra 
we derived the fundamental stellar parameters of $\rho$ Oph A and further constrained its age. 
We conclude that $\rho$\, Oph A provides strong evidence for the presence of active X-ray emitting regions 
on young magnetized early type stars.
}
\titlerunning{Magnetism of \rhoOph~A}
\authorrunning{I. Pillitteri et al.}
\maketitle
\section{Introduction}\label{sec:introduction}
X-ray emission is a common feature among massive stars of spectral types 
O through early B \citep{Oskinova2016}. 
In single O-stars, a soft X-ray 
emission is thought to be generated by intrinsic shocks in the stellar winds 
\citep{Owocki1988,Feldmeier1997a,Feldmeier1997b}. 
In binary systems, large-scale shocks associated with wind-wind collisions can manifest 
themselves in additional X-ray emission \citep{Stevens1992}.
In strongly magnetized massive early type stars, winds can be channeled and collide at high Mach 
numbers generating hard X-ray emission  \citep{Babel97,udDoula2002}. 

In this framework, the generation of X-rays in stellar winds is less probable in B type stars 
because of their weaker winds compared to those from O and Wolf-Rayet stars \citep{Prinja1989}.
Early B stars mark the transition from strong winds / X-ray emitters to weak winds stars / X-ray
dark stars toward the rise of coronal X-rays in late A and F type stars. 
The rate of detection in X-rays among B stars falls to about 50\% 
\citep{Wolk2011,Naze2011,Berghoefer1997}.  
Little is known about wind-wind collision in binaries with B-type non-supergiant companions. 
Recently, Ignace et al. (2017, ApJ submitted) showed that even in close B0V-type binaries 
the enhanced X-ray emission due to the wind-wind collision is not observed.
Only in a a few non super-giant B-stars, hard X-ray emission is detected and represents a signature 
of strong magnetic fields \citep{Petit2013}. 
In this context, we report the detection of magnetic field in the B2, X-ray bright star \rhoOph~A. 

Rho Ophiuchi designates a multiple system of B2 to B5 stars at a distance from the Sun of about 110 pc.
In particular, \rhoOph~A is a B2 star in a binary system with another B2 star separated by about 300 AU. 
The age of this star is estimated  in the 7-10 Myr range \citep{Pillitteri2016b}, 
as it is coeval of a group of low mass  disk-less stars that surround it. 
\rhoOph\ is at the center of a small cluster born from the first burst of star formation in the Rho 
Ophiuchi cloud.
We have observed \rhoOph\ with the X-ray telescope XMM-Newton in 2013 (50 ks) discovering that it emits 
X-rays \citep{Pillitteri2014c}. In 2016 monitoring observations for another 140 ks \citep{Pillitteri2017} 
confirmed that \rhoOph~A emits X-rays in a peculiar way not previously seen in any other B-type star. 
We observed two increases of the X-ray flux, with the second event being the strongest and, likely, a flare. 
The time elapsed between the two events is approximately 1.2 days; the same interval corresponds
to the rotational period of the star, and it allows to phase-fold the data of 2013 and 2016 observations. 
In \citet{Pillitteri2016} we hypothisezed that an active spot of magnetic origin is the cause of 
the periodic increase of X-ray flux. 
Here we present the detection of magnetic field in \rhoOph~A in spectra acquired  
with FORS2 at ESO-VLT. The structure of the paper is the following: in \ref{sec:observations}
we present the data, their analysis and our results, in  \ref{sec:conclusions} we discuss the results 
and present our conclusions. 

\section{Observations, data reduction, and results}\label{sec:observations}
The star \rhoOph~A (a.k.a. HR~6112) was observed 
with the FORS2 low-resolution spectropolarimeter \citep{appenzeller1998}, which is attached to the 
ESO/VLT UT1 (Antu) of the Paranal Observatory (Chile). The data were taken in two epochs (17th of 
July and 11th of August 2017), with a slit width of 0.4$\arcsec$, and the grism 600B. 
The grism and slit width lead to a resolving power of approximately 1700 and the spectra cover 
the 3250--6215\,\AA, which includes all Balmer lines, except H$\alpha$, and a number of He lines. 
For each epoch, the star was observed with a sequence of eight spectra obtained by rotating the quarter 
waveplate alternatively from $-$45$^\circ$ to $+$45$^\circ$ every second exposure (i.e. $-$45$^\circ$, 
$+$45$^\circ$, $+$45$^\circ$, $-$45$^\circ$, $-$45$^\circ$, $+$45$^\circ$, $+$45$^\circ$, 
$-$45$^\circ$). 
The exposure times and obtained signal-to-noise ratios (S/N) per pixel of 
Stokes $I$ calculated around 4950\,\AA\ are listed in Table~\ref{tab:observations}.
\begin{table}
\caption{Observing log and surface average longitudinal magnetic field values.  \label{tab:observations} }
\begin{center}
\resizebox{\columnwidth}{!}{
\begin{tabular}{l|cccc}
\hline \hline
Date & HJD$-$  & \# of  & Exp. & S/N  \\
     & 2450000 & frames & time &      \\
\hline
17/07/2017 & 57951.03705 & 8 & 3.0 & 2357  \\
11/08/2017 & 57976.06673 & 8 & 8.0 & 3094  \\
\hline \hline
Date & \bz\          & \nz\          & \bz\    & \nz\        \\ 
     & \multicolumn{2}{c}{Hydrogen} & \multicolumn{2}{c}{All} \\
\hline
17/07/2017 &   $-283\pm$107(2.6) &    37$\pm$96(0.4) &  $-128\pm$68(1.9) &     5$\pm$62(0.1) \\
11/08/2017 &   $ 485\pm$84(5.7)  & $-$47$\pm$86(0.5) & 404$\pm$55(7.3)   & $-$30$\pm$58(0.5) \\
\hline
\end{tabular}
}
\end{center}
\tablefoot{The heliocentric Julian date shown in column two is that of the beginning of 
the sequence of exposures. Column three gives the number of frames obtained during each night of observation, 
while column four shows the exposure time, in seconds, of each frame. 
Column five gives the S/N per pixel of Stokes $I$ calculated at about 
4950\,\AA\ over a wavelength range of 100\,\AA. 
\bz\ values (G) are obtained from the spectral regions covered by 
the hydrogen lines obtained from the Stokes $V$ and $N$ parameter spectrum, respectively. 
The value in parenthesis gives the detection level { (e.g. \bz/\sbz).} 
The same is reported under columns {\em All}, when using the whole spectrum. 
The uncertainties are that scaled by the $\chi^2$ of the linear fit 
\citep[see Sect.~3.4 of][]{bagnulo2012}.}
\end{table}

The FORS2 data were reduced and analysed using the pipeline thoroughly described in 
\citet{fossati2015}, which is based on the algorithms and recommendations given by 
\citet{bagnulo2012,bagnulo2013}. The surface average longitudinal magnetic field \bz\ was derived 
using the method first described in \citet{bagnulo2002}, which is based on extracting the slope of 
the linear regression of $V(\lambda)/I(\lambda)$ versus the quantity 
$-g_{\rm eff}C_z\frac{\lambda^2}{I(\lambda)}\frac{{\rm d}I(\lambda)}{{\rm d}\lambda}$. 
The quantities $V(\lambda)$ and $I(\lambda)$ are the Stokes $V$ and $I$ profiles, respectively, 
$g_{\rm eff}$ is the effective Land\'e factor, which was set to 1.25 except for the region of the 
hydrogen Balmer lines where $g_{\rm eff}$ was set to 1.0 \citep{bagnulo2012}, 
and $C_z$ is equal to 4.67$\times$10$^{-13}$\,\AA$^{-1}$G$^{-1}$. 
We validated the employed effective Land\'e factor of 1.25 by extracting from the VALD database 
\citep{ryabchikova2015} the information relative to the lines covered by the FORS2 spectrum. 
We further calculated the diagnosing null profile $N$, and hence \nz\ \citep[see][]{bagnulo2009}.
\citet{bagnulo2012} present a detailed discussion of the physical limitations of this technique. 
The \bz\ and \nz\ values were calculated using either the hydrogen lines or the whole spectrum.

\begin{figure*}[]
\centering
\resizebox{0.9\textwidth}{!}{
\includegraphics{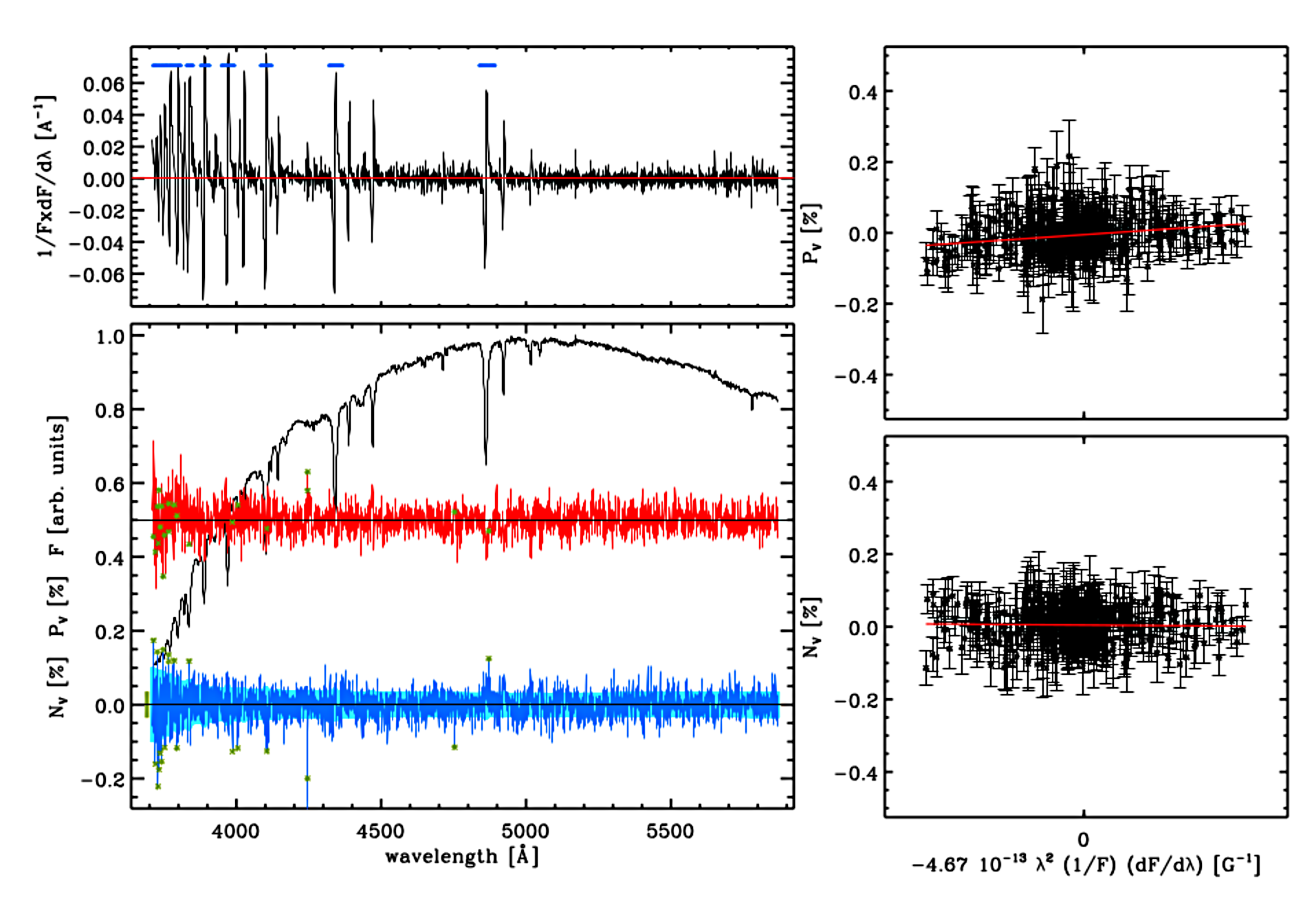}
}
\caption{ 
Top left panel: derivative of Stokes $I$. The regions used to calculate the magnetic field are 
marked by a thick blue line close to the top of the panel. Bottom left panel: the top profile shows 
Stokes $I$ arbitrarily normalised to the highest value, the middle red profile shows Stokes $V$ 
(in \%) rigidly shifted upwards by 0.5\% for visualisation reasons, while the bottom blue profile 
shows the spectrum of the $N$ parameter (in \%). The green asterisks mark the points that were 
removed by the sigma-clipping algorithm. The pale blue strip drawn underneath the $N$ profile 
shows the uncertainty associated with each spectral point. The thick green bar on the left side 
of the spectrum of the $N$ parameter shows the standard deviation of the $N$ profile. 
Top right panel: linear fit used to determine the magnetic field value using Stokes $V$ (i.e., \bz). 
The red solid line shows the best fit. 
Bottom right panel: same as the bottom left panel, but for the null profile (i.e., \nz). 
\label{fig:result}
}
\end{figure*}

\citet{bagnulo2012} showed that a magnetic field can be considered to be safely detected when 
\bz\,$>$5\,\sbz. Table~\ref{tab:observations} summarizes our results and shows that of the two sets of 
observations, the one obtained in August 2017 led to a definite magnetic field detection at a 
significance level $\ge6\sigma$. The spectra obtained in July 2017 led to a non-detection of 
the magnetic field. The reasons of such non-detection are discussed in Sect. \ref{sec:conclusions}.
The results of the analysis conducted using the hydrogen lines on the second data-set are visually 
shown in Fig.~\ref{fig:result}. We consistently found non-detection from the null profile 
(i.e., \nz\ consistent with zero).
\begin{table}
\caption{Stellar parameters determined for $\rho$\,Oph\,A.  \label{tab:params} }
\begin{center}                     
\resizebox{\columnwidth}{!}{
\begin{tabular}{lc|lc|lc}
\hline
\hline
\Teff\  [K]     & 22000$\pm$1000  	& 	M [\M]          & $8.2^{+0.8}_{-0.7}$   & $ 12 + \log \rm (C/H) $  & 7.7$\pm$0.2  \\
\logg  [cgs]    &   4.0$\pm$0.1 	&	R [\R]          & $4.5^{+0.6}_{-0.6}$   & $ 12 + \log \rm (N/H) $  & 7.5$\pm$0.2  \\
\vmic\  [\kms]  &   $2\pm1$   		&	$\tau$ [Myr]    & $15.3^{+4.2}_{-4.4}$  & $ 12 + \log \rm (O/H) $  & 8.8$\pm$0.2  \\
\vsini\  [\kms] &   $206_{-62.6}^{+27.3}$ & 	\logl           & $3.61^{+0.17}_{-0.16}$& $ 12 + \log \rm (Si/H)$  & 7.6$\pm$0.2  \\
\vmac\  [\kms]  &   $80.4_{-75.1}^{+125.7}$ &                   & 			& $ 12 + \log \rm (Mg/H)$  & 7.4$\pm$0.2  \\
\hline
\hline
\end{tabular}
}
\end{center}                     
\tablefoot{Uncertainties are 1$\sigma$-values. The \vsini\ value  is  obtained from the analysis of 
the \ion{Si}{iii} 4567\,\AA\ line with the FT+GOF method in \citet{Simon2014}. 
The stellar parameters mass, radius, age, and luminosity are obtained with {\sc bonnsai} \citep{Schneider2014} 
adopting the evolutionary tracks by \citet{Brott2011}.}
\end{table}

We further analysed archival high-resolution (R$\approx$115000) spectra obtained with 
the HARPS spectrograph, 
attached to the ESO 3.6\,m telescope in LaSilla, Chile. The star was observed on the 30th of March 2007 
when five consecutive spectra were obtained. The spectra have a peak signal-to-noise ratio (S/N) per 
resolution element of about 300, except for the third spectrum which we discarded because of its 
significantly lower quality (S/N\,$\approx$\,90). We retrieved the reduced, not normalised spectra 
from the ESO archive and, after looking for radial velocity variation, which we did not find, 
we co-added the four spectra. We finally obtained a spectrum with a peak S/N of about 600.

We derived the stellar atmospheric parameters using a grid of synthetic stellar atmosphere models 
built using the atmosphere/line formation code {\sc fastwind} \citep{Santolaya1997,Puls2005,Rivero2012}. 
The code takes into account non-local thermo-dynamical equilibrium effects in spherical symmetry with 
an explicit treatment of the stellar wind. The stellar grid is designed to model late O- and B-type 
stars with effective temperatures (\Teff) ranging from 12000 to 34000\,K in 1000\,K steps and surface 
gravity (\logg) between 2.0 and 4.4 in steps of 0.1\,dex. The helium abundance was kept fix to 
the solar value. Explicit atomic models for \ion{H}{i}, \ion{He}{i,ii}, \ion{N}{ii,iii}, \ion{O}{ii,iii}, 
\ion{C}{ii,iii}, \ion{Si}{ii,iii,iv}, and \ion{Mg}{ii} are included in the determinations of the 
atmospheric parameters. The other chemical species are treated in an implicit way to account for 
blanketing/blocking effects. 

The analysis is performed in two steps. First several key spectral lines are simultaneously compared with 
the grid looking for the set of stellar parameters that best reproduces the spectrum following the routines 
described in \citet{Castro2012} (see also \citealp{Lefever2010}). 
In a second step, new synthetic sub-grids are build for the star
 with chemical abundances spanning $\pm$2.00\,dex around the cosmic abundance standard in the solar neighborhood 
\citep{Nieva2012}, in steps of 0.2\,dex. The best combination of abundances that reproduces the observations was 
found through an optimized $\chi^2$ genetic algorithm. We further inferred the stellar projected rotational 
velocity \vsini\ through the Fourier transform (FT) + goodness-of-fit (GOF) method applied to the 
\ion{Si}{iii} $\lambda$4552 line profile using the {\sc iacob-broad} code \citep{Simon2014}.
{ We obtain similar \vsini\ using both the Fourier transform (206 km/s) 
and the GOF (208 km/s) techniques.
The large \vsini\ do not allow the GOF to constrain any additional broadening
introduced by macro-turbulence. However, the \vsini+macro-turbulence degeneration has a secondary role in the stellar 
atmosphere quantitative analysis since similar profiles will also retrieve a similar match within the 
synthetic grid of models. The stellar profile seems determined by rotation, and macro-turbulence 
is not playing any significant role that can be distinguished under the rotation profile. 
} 
The results of the analysis are listed in Table~\ref{tab:params}. 
Although it leads to a slightly worse fit to the data, the code allowed also for a solution with a 
higher \Teff\ and \logg\ by 1000\,K and 0.1\,dex, respectively. 
A higher \Teff\ and \logg\ values would lead to a slight increase in the stellar mass and radius 
and decrease in age, which would then better fit the age of the \rhoOph\ group.
Despite the high data quality, the large abundance uncertainties are due to the difficulty of obtaining a 
secure normalisation of the spectrum due to the very broad lines. The stellar mass (M), radius (R), luminosity 
(\logl), and evolutionary age ($\tau$) in Table~\ref{tab:params} were derived from comparisons with 
stellar evolutionary tracks by \citet{Brott2011} using the {\sc bonnsai}\footnote{The BONNSAI web-service is 
available at {\tt www.astro.uni-bonn.de/stars/bonnsai}.} tool \citep{Schneider2014,Schneider2017}. For these 
estimations, we employed tracks which account for stellar rotation. We further derived the stellar parameters 
without considering rotation obtaining results which are within 1$\sigma$ of those listed in Table~\ref{tab:params}.
From the stellar radius and rotation period of 1.2\,d, we infer an equatorial rotational velocity of 190$\pm$25\,\kms, 
which is consistent within 1$\sigma$ with the measured \vsini\ value. This suggests that the star is viewed equator-on, 
hence the inclination angle of the stellar rotation axis is close to 90$^\circ$, rather than 45$^\circ$ as inferred
in \citet{Pillitteri2017}, the difference arising from an assumed radius of 8 R$_\odot$, 
and based on a rough estimate by \citet{vanBelle2012}.
Following the formalism of \citet{Petit2013}, with the stellar parameters listed in Table~\ref{tab:params}, 
a rotation period of 1.2\,days, and a lower limit on the dipolar magnetic field of 1.5\,kG, we derived a Keplerian 
co-rotation radius of 2.1 stellar radii and a lower limit on the Alfv\'en radius of 17.6 stellar radii. For these 
calculations, we further adopted a mass-loss rate of 1.4$\times$10$^{-10}$\,\M\,yr$^{-1}$ \citep{Krticka2014} 
and a  terminal velocity of 700\kms\ \citep{Prinja1989,Oskinova2011}. 
Following the results of \citet{Petit2013}, $\rho$\,Oph\,A should host 
a centrifugal magnetosphere, and the interplay between the stellar magnetic field and 
rotation rate  may be able to sustain a circumstellar disk. At the light of this prediction, 
we looked for emission in the H$\alpha$ line profile, but found no convincing evidence of it.
\section{Discussion and conclusions}\label{sec:conclusions}
We have detected magnetism in the B2 star \rhoOph~A during one of the two epochs of FORS2 observations. 
In detail, we estimated that the longitudinal component of the field is about \bz\,$\sim490\pm$80\,G. 
In the other set of observations we do not detect the magnetic field, this does not necessarily implies that the
magnetic field is zero.
The estimate of the magnetic field are averaged over the full disk of the star and 
as such it strongly depends on the configuration of the magnetic field and on the rotational phase at which we 
observe the star.  
From the variability in X-rays of \rhoOph~A, we inferred that a likely value of the stellar 
rotational period is $\sim1.205$ days \citep{Pillitteri2017} and that, having measured 
v$\sin i \sim 206$ km/s from optical spectra, the $i$ angle is about $90\deg$.
It is likely that the magnetic axis is misaligned with respect to the rotational axis of the star,
as commonly found in massive stars. This results in variations of the surface average longitudinal magnetic field \bz,
with a period equal to the rotation period. This would explain why the field was not detected in the first data set.
A more specific and systematic monitoring on a time scale of one day is thus needed to measure more 
precisely the variation of the magnetic field along the stellar rotational phases. 

The presence of a magnetic field is in strong agreement with the interpretation of the X-ray light
curve of \rhoOph~A given in \citet{Pillitteri2017}. In particular, an active spot, which is the source of the 
periodic X-ray brightenings and transits in about 30 ks (8.3 hr), 
would be created by a strong local magnetic field which average intensity is almost
500 G. Such number is also consistent with the estimate of the minimum magnetic field strength ($B>300$ G) 
required to constrain the hot plasma flaring during the second event. The estimate of the magnetic field in
the flaring loop was obtained applying the diagnostics of \citet{Reale2007} to the rise and the decay of the
temperature and emission measure of flares in stellar coronae.
After obtaining ephemeris of the passage of the spot from the X-ray light curve, we checked 
that during the FORS2 observations the spot  was not on view. 
{ The interval occurred between the \xmm\ and FORS2 observations is of order of 500 days.
The uncertainty of the ephemeris of the spot is of order of 0.001 days from the X-ray analysis,
this uncertainty can result in a transit timing off by about 0.5 day or about $\sim0.4$ in phase. 
Hence a proper monitoring in optical band and X-rays is needed to better understand any 
modulation of the magnetic field as a function of the stellar rotation. } 

Two spotted and magnetic early B stars in the young open cluster of NGC~2264 have been discovered 
by \citet{Fossati2014}, namely HD~47887 and HD~47777. For HD~47887, there is a X-ray detection 
in \chandra\ with a flux of $6.03\times10^{-15}$ \fxu, at a distance of 800 pc this means a X-ray luminosity of 
$\sim4.2\times10^{29}$ \lxu, which is lower than that observed on average in \rhoOph~A.
For HD~47777, there are two X-ray detections in two different \xmm\ observations, both within $1\arcsec$ from the optical
position of the star, with X-ray fluxes of $5.3\times10^{-14}$ \fxu and $9.03\times10{-14}$ \fxu, 
that means X-ray luminosities of $3.7\times10^{30}$ \lxu and $6.4\times10^{30}$ \lxu,
respectively. However, \rhoOph~A shows the best example of modulated X-ray emission likely due to intrinsic 
magnetism, with X-ray luminosity varying in the range $2.5-12.3\times10^{30}$ \lxu.

To our best knowledge, \rhoOph~A offers the first detection of X-ray emission associated
with a localized area on the stellar surface of a massive star. 
Previously, large-amplitude, periodic X-ray emission was reported for some magnetic O type
stars, such as $\theta^1$ Ori C \citep{Gagne1997} and NGC 1624-2
\citep{Petit2015}. In these large magnetospheres, X-ray
variability is likely associated with the axisymmetric distribution of confined stellar
wind material. Previous dedicated attempts to detect X-ray variability in B-type stars
were not successful. For example, no X-ray variability associated with rotation was detected in
the magnetic B-type star $\tau$ Sco \citep{Ignace2010}.
It is especially interesting that X-ray observations suggest a flaring activity 
associated with the magnetic spot. Albeit previously X-ray flares from B-type stars were discussed
in the literature \citep{Mullan2009}, $\rho$~Oph~A presents an
so far a unique case of periodic X-ray activity that might be associated with
an active region on the stellar surface.
In \citet{Pillitteri2017} we also discuss the hypothesis that a low mass companion is the emitter of X-rays. 
Such a companion would orbit in 1.2 days and would be partially obscured by the primary. 
In this case the magnetic field detection would be associated with this object and the strength of the magnetic 
field would be much higher as the value is an average over the full disk of the primary. 
The extant optical spectra of \rhoOph~A do not present spectral lines  
associated to such companion. A dedicated spectroscopic monitoring of \rhoOph~A is thus needed 
to detect it (if at all present) from, e.g.,  RV measurements although the very broad lines 
of \rhoOph~A hamper this search.   

\begin{acknowledgements}
IP acknowledges support from INAF, ASI and the ARIEL consortium.
LMO acknowledges support by DLR grant 50OR1302 and partial support by the Russian Government 
Program of Competitive Growth of Kazan Federal University.
S.J.W. was supported by NASA contract NAS8-03060 (Chandra X-ray Center)
\end{acknowledgements}
%

\end{document}